# Persistent Hierarchy in Contemporary International Collaboration

Lili Miao[1], Vincent Larivière[2,3], Byungkyu Lee[4], Yong-Yeol Ahn[1,*], Cassidy R. Sugimoto[5,*]

[1]Center for Complex Networks and Systems Research, Luddy School of Informatics, Computing, and Engineering, Indiana University Bloomington, Bloomington, IN, USA.
[2]École de bibliothéconomie et des sciences de l'information, Université de Montréal, Montréal, Québec, Canada.
[3]Observatoire des sciences et des technologies, Université du Québec à Montréal, Montréal, Québec, Canada.
[4]Department of Sociology, New York University, New York, NY, USA.
[5]School of Public Policy, Georgia Institute of Technology, Atlanta, GA, USA.
*Email Addresses: yyahn@iu.edu (Yong-Yeol Ahn), sugimoto@gatech.edu (Cassidy R. Sugimoto)

**Abstract**
Science is increasingly global, with international collaboration playing a crucial role in advancing scientific development and knowledge exchange across borders. However, the processes that regulate how scientific labor is distributed among countries remain underexplored, leading to challenges in ensuring both effective collaboration and equitable participation across diverse scientific communities. Here, we leverage three million internationally coauthored publications produced by countries worldwide to examine the division of scientific labor in international collaboration, identify the factors that shape this distribution, and assess its broader consequences. Our findings uncover a persistent hierarchical structure in international collaboration, with researchers from scientifically advanced countries occupy dominant roles, while those from less-developed countries are relegated to supportive roles, even after controlling for other influential factors. This hierarchy is also reflected in the research content, as countries with lower scientific capacity tend to participate in international collaborations that deviate from their domestic science. By analyzing the labor division within international collaborations, we demonstrate that researchers from less-developed countries face systematic disadvantages, which not only limit their contributions to the global scientific community but also prevent them from fully benefiting from international collaborations.

**Introduction**

International collaboration has increasingly become the cornerstone of scientific progress, particularly in addressing complex global challenges such as vaccine development for pandemics and combating climate change—problems that require contributions from researchers worldwide[1–3]. These collaborations not only accelerate scientific discovery by uniting diverse expertise and resources but also facilitate the flow of knowledge across borders, strengthening national scientific capacities[4,5]. International collaboration inherently involves the distribution of scientific labor across countries, with researchers contributing their unique strengths, specializations, and infrastructure to make these efforts possible. However, despite the rising importance of international collaboration, little research has examined how the scientific labor is distributed across countries and what factors influence the distribution of scientific labor.

Historically, international collaboration has roots in the exploitative practices of Western colonial powers[6,7]. During the colonial period, scientists from industrialized nations were often dispatched to colonized regions to assess and extract natural resources[6]. They collected data and samples in those regions, returned to their home countries, analyzed the findings, and published results with minimal involvement from local researchers[7,8]. Unfortunately, this practice persists into modern international collaborations.

Existing studies have revealed that researchers from developed countries are more likely to lead in formulating research questions, designing studies, and supervising projects[9–11]. In contrast, researchers from less-developed countries are often perceived as unqualified to lead, relegating them to subordinate roles such as providing specimens or conducting fieldwork[9–12]. This entrenched division of labor disproportionately benefits researchers from developed countries, while the contributions of those from less-developed regions are undervalued[13]. For instance, researchers from less-developed countries are frequently underrepresented in prominent authorship positions[14,15] or even excluded from co-authorship entirely[16,17]. This marginalization has significant negative consequences for their career advancement in academic systems, where publications are a key metric for promotion and reputation buiding[18]. Moreover, this labor exploitation harms local societies in less-developed countries. Without sovereignty over international collaborations, researchers in these regions often see their scientific priorities sidelined[19], leading to research agendas that are misaligned with local societal needs[20].

However, this traditional model of international collaboration, characterized by power asymmetry and inequitable labor distribution, is increasingly being challenged by the rise of emerging nations—particularly China[21–24]. Over the past few decades, China has become a major scientific powerhouse, rivaling the United States in global influence[25]. China has expanded its reach, especially to countries in the global South, positioning itself as a more moderate partner and seeking to build alliances aimed at combating the "exploitation and oppression" rooted in the colonial era[25].

Given the critical role of international collaboration, alongside the adverse effects of power asymmetry rooted in colonial science and the evolving global scientific landscape, there is a pressing need to examine how international collaborations are conducted and how scientific labor is distributed across countries. In this study, we analyze publications records from 201 countries to investigate the distribution of labor in international collaborations. Each internationally coauthored publication is viewed as a product of collaborative effort, with the labor distribution inferred from authorship order. Previous research has demonstrated that authorship order in multiauthor papers signals the roles of contributions[26,27] and serve as a key metric in academic evaluations[28]. We analyze division of labor by examining the relationship between authors' nationalities and their positions on the authorship list. Since the first and last authors are typically credited with the most significant contributions to the research[26,29] and receive the highest recognition in evaluation systems[28], we classify authorship positions into leading roles (first and last authors) and supportive roles (middle authors). We then investigate whether authors from less-developed countries are disproportionately relegated to supportive roles, while controlling for other relevant factors, and explore the potential consequences of this unequal labor division.

**Results**

As the global scientific development and its integration march on, the number of authors participating in international collaborations has been steadily increasing, with researchers from scientifically advanced countries still comprising the largest proportion (see Methods, Fig. 1a-b). Meanwhile, there has been a substantial increase in the involvement of researchers from scientifically proficient countries in international collaborations, accompanied by a corresponding rise in the share (see Fig.1a). However, while the number of researchers from scientifically developing and lagging countries participating in international collaborations has also grown, their proportion remains modest (see Fig.1a). This trend holds across various authorship roles, with authors from scientifically advanced countries dominating the largest share across all positions, researchers from scientifically proficient countries have notably risen in all authorship positions, particularly as first authors, alongside a more modest contribution from researchers locate at developing and lagging countries (see Fig. 1b).

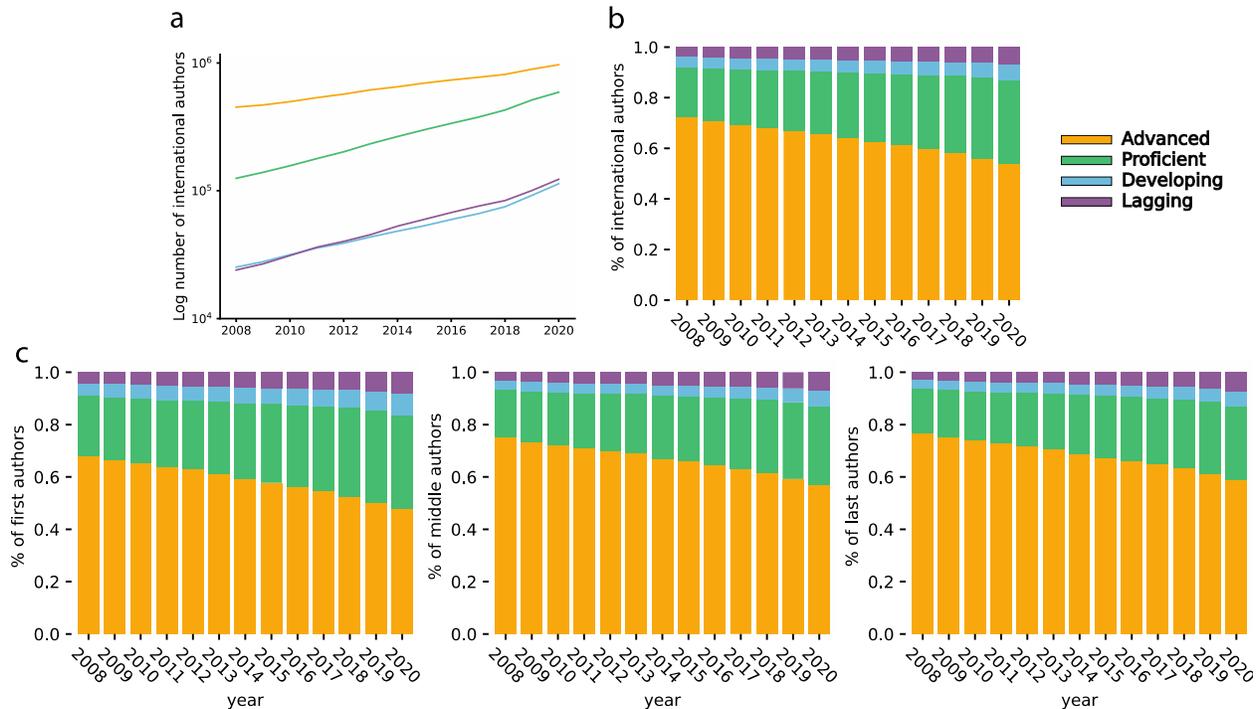

*Figure 1 **Increasing proportion of authors from scientifically proficient countries in the international scientific workforce, especially as first authors.** (a) Number of authors participating in international collaborations within each scientific capacity group over years. (b) Proportion of international scientific workforce across different scientific capacity groups. (c) Proportion of authorships is occupied by authors from each scientific capacity group.*

Although a large proportion of authors in international collaborations come from scientifically advanced countries, the distribution of authorship positions is influenced by the size of a country's scientific workforce and its share of internationally coauthored publications. To account for the influence of country size and more accurately assess the tendency of researchers from different countries to take on various roles, we compare the observed occurrences of each authorship position for each country with the expected values (see Methods). This comparison reveals a hierarchical structure in the division of scientific labor.

Researchers from scientifically proficient countries and developing countries are more likely to occupy the first author position, suggesting that they often take on "labor-intensive" tasks such as conducting research and performing formal analyses[29]. In contrast, researchers from scientifically advanced countries are more likely to assume the last author position, indicating their roles in conceptualizing and supervising the project[29]. Notably, the last author role, often regarded as the most prestigious, is disproportionately occupied by researchers from advanced countries, while those from non-advanced countries are constantly underrepresented in this position (see Fig. 2a-b). Although there is a growing trend of researchers from scientifically proficient countries assuming the last authorship, they remain underrepresented in this role compared to advanced countries. Researchers from scientifically lagging countries, however, are disproportionately underrepresented in both first and last authorship positions, while being overrepresented as middle authors. This suggests that they are more likely to contribute to supportive tasks, such as data curation, investigation, or software development, rather than taking on leadership roles in international collaborations[29]. Despite a large number of middle authors being based in advanced countries, when accounting for the size of the scientific workforce, researchers from scientifically advanced countries are less likely than expected to occupy the middle author role in international collaborations (see Fig. 2).

To ensure that these results are not driven by collaborations between countries with similar scientific capacities, we conducted the same analysis, excluding publications coauthored by countries within the same capacity group. The results remain consistent, with an even more pronounced divide between scientifically advanced and non-advanced countries (see Fig. S1). When researchers from scientifically advanced countries collaborate with those from different scientific capacity groups, they are significantly more likely to occupy the last author position (see Fig. S1).

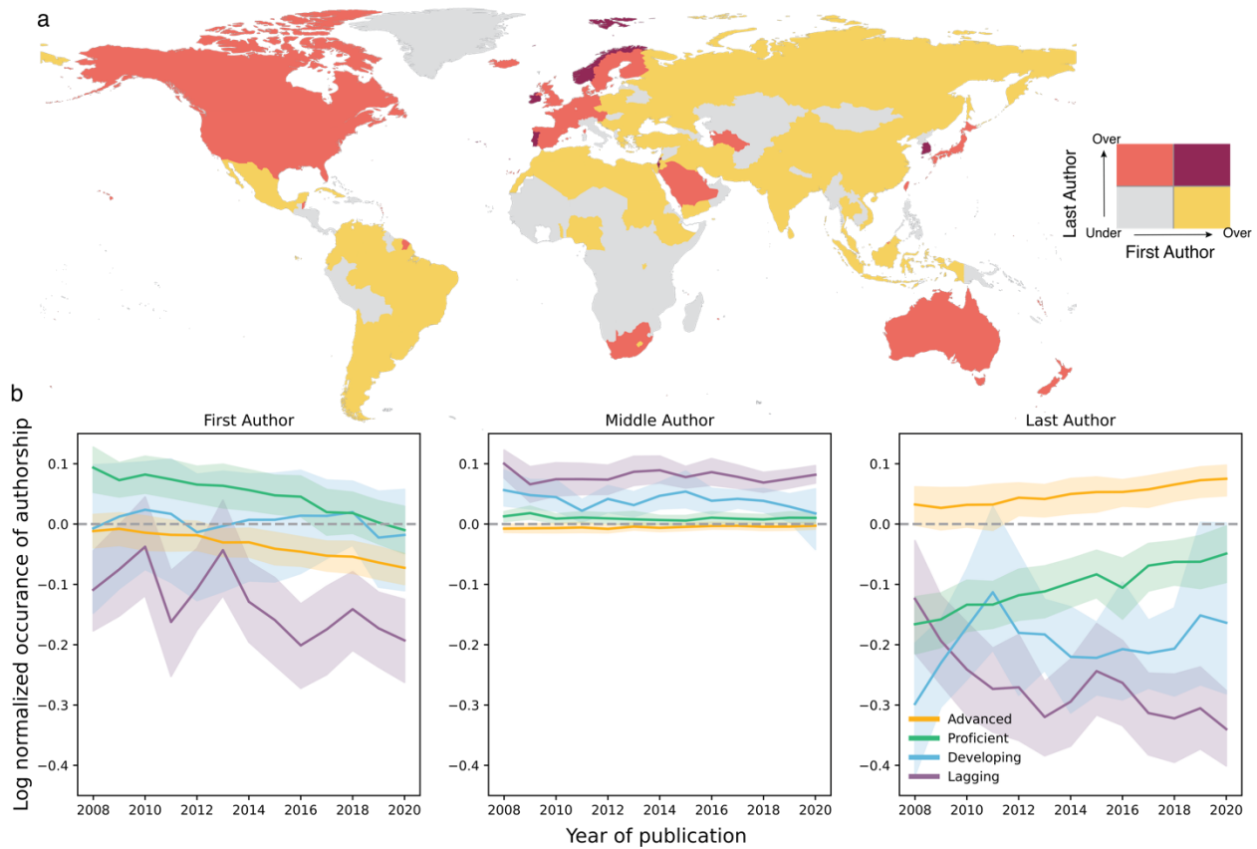

*Figure 2 **The authorship order in internationally collaborated papers follows a hierarchcial structure.** (a) Countries are color-coded to indicate their representation in first and last authorship positions. Dark red represents that researchers from the focal country are overrepresented in both first and last authorship. Light red indicates researchers from the focal country are underrepresented in first authorship while overrepresented in last authorship. Yellow represents researchers from the country are overrepresented in first authorship but underrepresented in last authorship. The gray area represents researchers from the country are underrepresented in both first and last authorship. (b) Temporal trend. The occurrence of authorships by researchers from different countries is normalized against the expected number (see Methods). The Y-axis shows the logarithm of the normalized value. A positive value indicates that researchers from that country appear more frequently than expected, while a negative value indicates they appear less frequently than expected. The shaded area indicates the confidence interval, which is derived from aggregating countries in the same scientific group.*

While the normalization results reveal that researchers' role and importance in international collaboration is associated with their affiliation countries, the process of determining authorship order is simultaneously influenced by various factors, including the scientific capacity of authors[26,29] and the author's gender[30] (see Data and Methods, see Fig. 3a). To identify the impact of nationality on authorship while controlling for the influence of other relevant factors, we apply a paper-level fixed-effects regression model (see Data and Methods). The regression results confirm that researchers from non-advanced countries tend to be systematically underrepresented in the role of the last author; instead, they tend to assume the role of the first author and middle author (see Table 1). Specifically, women are more likely to serve as the first author than men, while less likely to serve as the last author. Compared to researchers from scientifically advanced countries, those from non-advanced countries are more likely to be the first author, instead of the last author, with researchers from scientifically proficient countries exhibiting the highest effect size. Meanwhile, researchers from non-advanced countries are also more likely to play the role of middle author, with the researchers from scientifically lagging countries have the highest effect size. While funding has a positive impact on increasing the likelihood of assuming leading roles as the first and last author, funded researchers from lagging countries still have a lower chance of serving as the last author compared to unfunded researchers from scientifically advanced countries (see Fig. 3b).

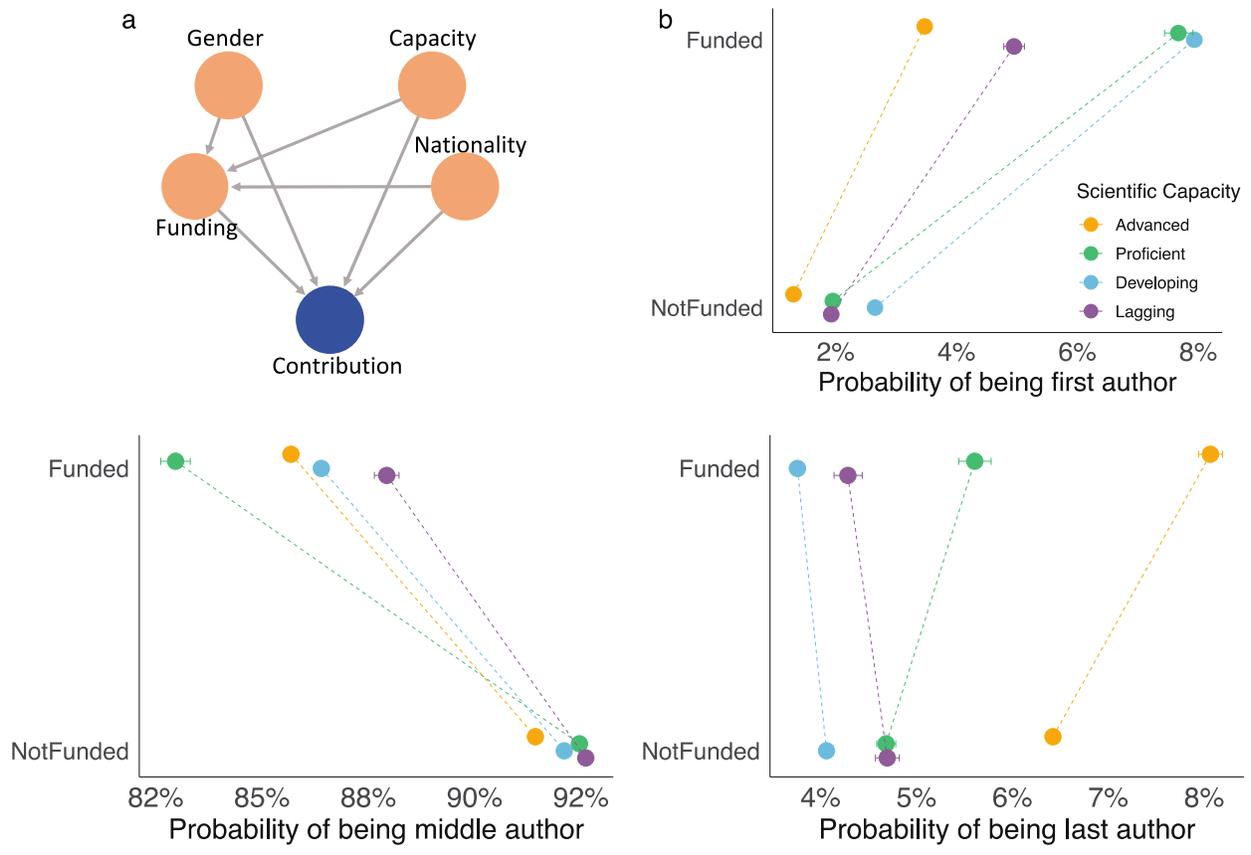

Figure 3 **Researchers from scientifically non-advanced countries are less likely to assume the role of the last author, even when they provide funding for the research**. (a) Causal diagram for the regression model. (b)The marginal effect of funding on playing the role of the first author, middle author, and the last author.

Table 1 Fixed-effects regression results at the paper level with authorship order as the dependent variable.

|  | First Author | | | Middle Author | | | Last Author | | |
|---|---|---|---|---|---|---|---|---|---|
|  | (1) | (2) | (3) | (1) | (2) | (3) | (1) | (2) | (3) |
| Male | -0.247*** | -0.212*** | -0.213*** | 0.022*** | -0.004 | -0.004 | 0.235*** | 0.244*** | 0.243*** |
|  | (0.003) | (0.004) | (0.004) | (0.002) | (0.003) | (0.003) | (0.003) | (0.005) | (0.005) |
| Gender(Unknown) | -0.282*** | -0.254*** | -0.254*** | 0.021*** | 0.009* | 0.009* | 0.211*** | 0.195*** | 0.194*** |
|  | (0.003) | (0.006) | (0.006) | (0.003) | (0.004) | (0.004) | (0.004) | (0.006) | (0.006) |
| Log no.pub | -0.197*** | -0.201*** | -0.201*** | -0.129*** | -0.137*** | -0.137*** | 0.443*** | 0.454*** | 0.455*** |
|  | (0.001) | (0.001) | (0.001) | (0.000) | (0.001) | (0.001) | (0.001) | (0.001) | (0.001) |
| Proficient | 0.627*** | 0.707*** | 0.412*** | -0.190*** | -0.101*** | 0.140*** | -0.279*** | -0.355*** | -0.334*** |
|  | (0.002) | (0.004) | (0.010) | (0.002) | (0.004) | (0.008) | (0.003) | (0.005) | (0.010) |
| Developing | 0.556*** | 0.839*** | 0.724*** | 0.092*** | 0.055*** | 0.090*** | -0.638*** | -0.644*** | -0.483*** |
|  | (0.005) | (0.010) | (0.014) | (0.004) | (0.008) | (0.011) | (0.006) | (0.011) | (0.016) |
| Lagging | 0.251*** | 0.438*** | 0.397*** | 0.164*** | 0.152*** | 0.162*** | -0.449*** | -0.503*** | -0.332*** |
|  | (0.004) | (0.009) | (0.013) | (0.004) | (0.007) | (0.010) | (0.005) | (0.010) | (0.013) |
| Funded |  | 1.155*** | 1.003*** |  | -0.681*** | -0.578*** |  | 0.170*** | 0.245*** |
|  |  | (0.005) | (0.008) |  | (0.004) | (0.006) |  | (0.005) | (0.007) |
| Proficient × Funded |  |  | 0.425*** |  |  | -0.346*** |  |  | -0.056*** |
|  |  |  | (0.014) |  |  | (0.011) |  |  | (0.014) |
| Developing × Funded |  |  | 0.149*** |  |  | -0.031+ |  |  | -0.326*** |
|  |  |  | (0.022) |  |  | (0.017) |  |  | (0.022) |
| Lagging × Funded |  |  | -0.028 |  |  | 0.036* |  |  | -0.341*** |
|  |  |  | (0.019) |  |  | (0.016) |  |  | (0.020) |
| Num.Obs. | 12164877 | 4640948 | 4640948 | 12035527 | 4603678 | 4603678 | 12164877 | 4640948 | 4640948 |
| R2 | 0.086 | 0.104 | 0.105 | 0.097 | 0.100 | 0.101 | 0.130 | 0.133 | 0.133 |
| R2 Adj. | -0.271 | -0.252 | -0.252 | -0.147 | -0.144 | -0.144 | -0.226 | -0.223 | -0.223 |

+ p < 0.1, * p < 0.05, ** p < 0.01, *** p < 0.001

Although authorship position has been widely used to infer an author's contribution[27,29] due to the absence of large-scale detailed labor division data, it fundamentally differs from an author's actual contribution[30]. This discrepancy suggests that the previous results may conflate true labor division with authorship assignment conventions. To more accurately capture the labor division in international collaboration, we leverage a small-scale dataset from PLOS journals containing self-disclosed credit contributions of authors and examine the relationship between authors' country of affiliations and their contributions in international collaborations (see Data and Methods). The result first reveals, on average, researchers from scientifically advanced countries contribute a larger share of credits compared to those from non-advanced countries (see Data and Methods, Fig. 4a). By comparing the complementarity of credits contributed by researchers from countries with varying levels of scientific capacity (see Data and Methods), the result suggests that researchers from scientifically non-advanced countries tend to perform relatively different tasks compared to their counterparts from scientifically advanced countries during collaborations (see Fig. 4b). Specifically, a significant proportion of tasks are exclusively performed by researchers from either scientifically advanced or non-advanced countries. However, this proportion decreases when researchers from scientifically lagging countries are involved, indicating that tasks performed by these researchers show greater overlap with those performed by their coauthors than the tasks carried out by researchers from scientifically advanced countries (see Fig. 4b).

To further understand task specialization among researchers, we compare the observed occurrences of each type of credit by country with the expected values. This comparison again reveals a clear divide between researchers from scientifically advanced countries and those from non-advanced countries. Contributions that are disproportionately performed by researchers from advanced—such as writing, conceptualization, methodology—are consistently underrepresented among researchers from non-advanced countries. In contrast, contributions that are underrepresented by researchers from advanced countries, including data curation, resources, investigation and project administration, are all overrepresented by those from non-advanced countries (see Fig. 4c).

Similar to authorship assignment, researchers' contributions to studies are influenced by various factors, including gender and seniority[29]. Therefore, to clearly reveal the relationship between authors' country of affiliations and their contributions while controlling for these influences, we apply the aforementioned paper level fixed-effects regression model with credit contribution as the outcome variable (see Fig. 4d). Given that an author's contribution to a paper cannot be fully captured by any single contribution type, we classify credit contributions into two groups—strategic and supportive—in order to distinguish the central roles that drive the intellectual and strategic direction of the research from those that provide more supportive functions (see Data and Methods). The regression result reveals that, after controlling for confounding factors such as gender and the scientific capacity of researchers, researchers from scientifically developing and lagging countries are still significantly less likely to contribute to strategic roles (see Table 2). Specifically, compared to researchers from advanced countries, those from scientifically developing and lagging countries have 37% and 46% lower odds of contributing to strategic roles in international collaborations, respectively (see Table 2). To further explore the relationship between country of affiliation, labor division and credit assignment in international collaborations, we expand the regression model to include whether the author performed strategic roles as an independent variable (see Fig. 4e). This allows us to investigate how an author's country of affiliation influences credit assignment in collaborations, even after accounting for the strategic contributions they made. The results reveal a consistent hierarchical structure in authorship order: researchers from scientifically proficient and developing countries are more likely to hold the position of first author than their counterparts from advanced countries, even after controlling for the strategic contributions made. In contrast, researchers from scientifically lagging countries are less likely to be listed as first authors as well as last authors. These researchers are often relegated to middle authorship roles, despite contributing at similar levels to their collaborators (see Table 2).

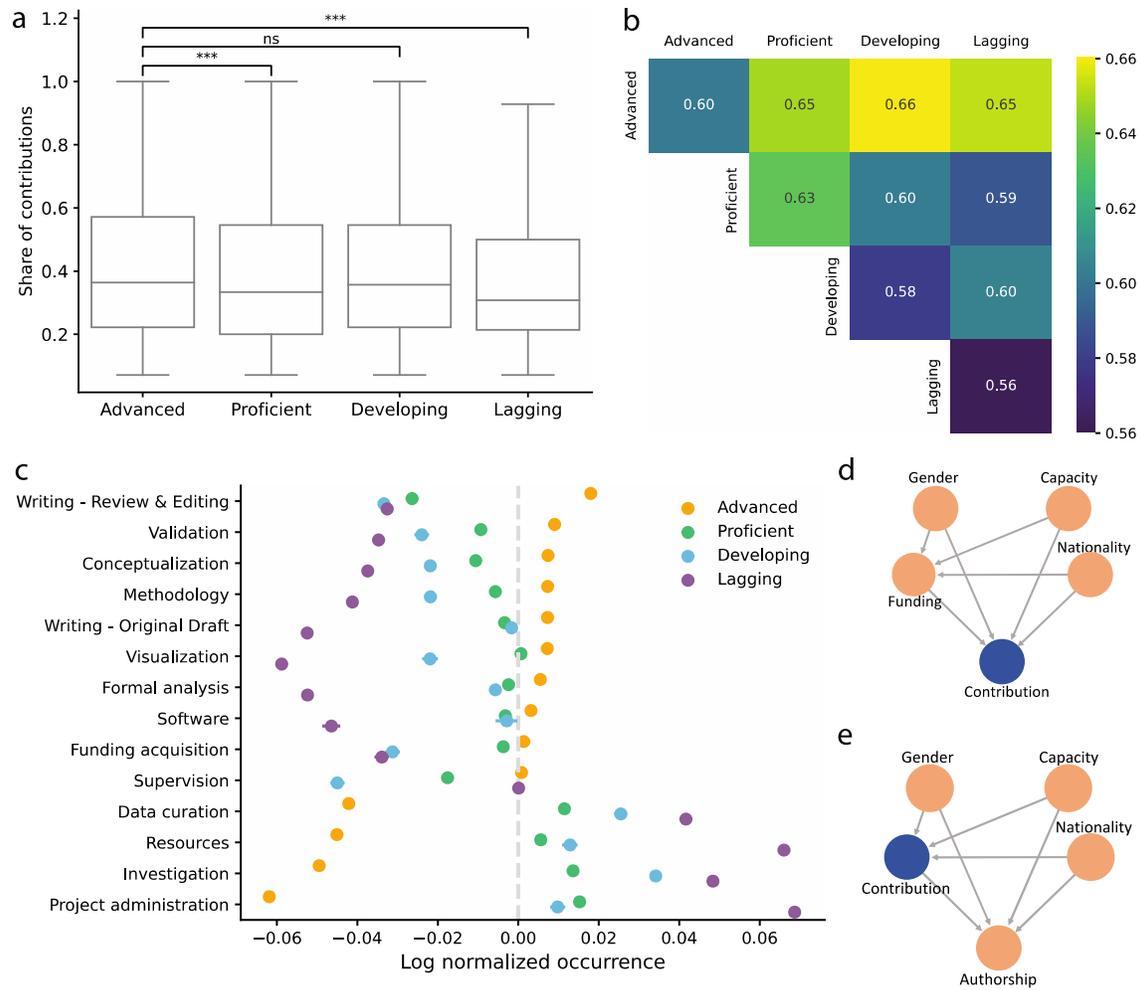

Figure 4 **Researchers from scientifically advanced countries are overrepresented in contributions related to intellectual work.** (a) The share of contributions made by researchers from countries with different levels of scientific capacity. The significance is calculated using a one-sided t-test for independent samples to decide whether the mean value of the scientifically advanced countries is significantly greater than the other countries. The significance levels are denoted as follows: ns indicates not significant, + indicates P<0.1, ∗ indicates P<0.05, ∗∗ indicates P<0.01, and ∗∗∗ indicates P<0.001. (b) The complementarity of contributions between researchers from countries with different scientific capacities. (c) Contributions made by researchers from different scientific groups are normalized by the expected value derived from randomly shuffling the contributions within each paper. A normalized occurrence greater than 0 indicates that researchers from the group are overrepresented in that contribution, while values less than 0 indicate underrepresentation. (d) Causal diagram with contribution as the outcome variable. (e) Causal diagram with contribution as the independent variable and authorship position as the dependent variable.

*Table 2 Fixed-effects regression results at the paper level considering both contribution and authorship order.*

|  | Strategic Role | First Author | Middle Author | Last Author |
|---|---|---|---|---|
| Male | 0.051 | -0.422*** | 0.220*** | 0.009 |
|  | (0.040) | (0.059) | (0.045) | (0.066) |
| Gender(Unknown) | -0.051 | -0.206* | 0.072 | -0.048 |
|  | (0.056) | (0.083) | (0.065) | (0.095) |
| Log no.pub | 0.325*** | -0.362*** | -0.104*** | 0.610*** |
|  | (0.013) | (0.016) | (0.012) | (0.022) |
| Proficient | 0.049 | 0.406*** | -0.244*** | 0.077 |
|  | (0.061) | (0.071) | (0.064) | (0.079) |
| Developing | -0.458*** | 0.350* | -0.108 | -0.016 |
|  | (0.130) | (0.149) | (0.122) | (0.163) |
| Lagging | -0.606*** | -0.350*** | 0.586*** | -0.605*** |
|  | (0.078) | (0.094) | (0.078) | (0.109) |
| Strategic Role |  | 2.197*** | -2.235*** | 1.912*** |
|  |  | (0.073) | (0.055) | (0.078) |
| Num.Obs. | 25379 | 25278 | 25278 | 25278 |
| R2 | 0.203 | 0.144 | 0.178 | 0.208 |
| R2 Adj. | 0.036 | -0.177 | -0.034 | -0.113 |

+ p < 0.1, * p < 0.05, ** p < 0.01, *** p < 0.001

The revealed labor division and authorship distribution confirm that the current structure of international collaboration remains entrenched in an asymmetric power dynamic, where researchers' contributions and the credit they receive are largely predetermined by their country of affiliation, with larger countries playing dominant roles in the system. This imbalance not only leads to labor exploitation of researchers from less advanced countries but also influences the research agenda of international collaborations. To investigate which countries have larger autonomy in shaping research agendas in international collaborations, we measure the topical similarity between domestic and international publications across countries. The result indicates an inverse relationship between a country's scientific capacity and the topical divergence between its domestic and international publications (see Fig. 4a). Specifically, countries with higher scientific capacity exhibit lower topical divergence between domestic and international publications, indicating that their researchers tend to work on similar topics regardless of whether they collaborate internationally or not (see Fig. 4a). In contrast, the topical divergence increases as scientific capacity declines, with the highest dissimilarity observed in scientifically lagging countries, suggesting the researchers from these countries work on relative different topics when they engage in international collaborations (see Fig. 4a).

Two potential explanations may account for the inverse relationship between national scientific capacity and topical divergence. First, the global research agenda appears to be driven primarily by scientifically advanced countries, with other countries aligning their research topics accordingly, irrespective of international collaborations. To test this hypothesis, we measure content similarity between each pair of countries using their domestic publications. Our findings reveal a negative relationship between a country's scientific capacity and its topical divergence from other countries (see Fig. 4b). Scientifically advanced countries exhibit the lowest topical divergence with other countries, while lagging countries show the highest topical divergence. Given that countries around the world tend to align their research agendas ore closely with those of scientifically advanced countries, this explains why countries with higher scientific capacity experience lower topical divergence in international collaborations.

The second explanation concerns the sovereignty in setting research priorities within international collaborations. Since leading authors typically have a greater influence over research direction than non-leading authors, and researchers from less developed countries are less likely to take on leading roles, we examine whether internationally coauthored publications with researchers from a given country in leading roles (either first or last author) exhibit higher topical similarity to their domestic publications. Our results indicate that internationally coauthored publications with researchers in leading roles show significantly higher similarity to their country's domestic publications compared to those where the researchers occupy middle authorship roles (see Fig. 4b). This pattern is consistent across countries with varying scientific capacities (see Fig. 4b). Consequently, given that researchers from scientifically lagging countries are more likely to occupy supportive roles, it follows that their international publications are less aligned with their national research agendas.

To examine whether the observed patterns could be attributed to publication sampling biases, due to differences in publication output across countries, we conduct a counterfactual experiment assuming that each country produced an equal number of publications in each discipline and then we replicate the previous analysis (see Data and Methods). The observed patterns remain consistent: content divergence increases as scientific capacity decreases (see SI). However, the range of divergences between groups narrows, suggesting that while the relationship between content similarity and scientific capacity is robust, the observed pattern is still influenced by the sampling bias inherent in the varying publication sizes. Therefore, we believe that the true group-level difference lies between the divergence observed in the empirical data and that of the counterfactual scenario. This pattern is also robust in the topical divergence of domestic publications and the divergence between leading and supportive publications (see SI).

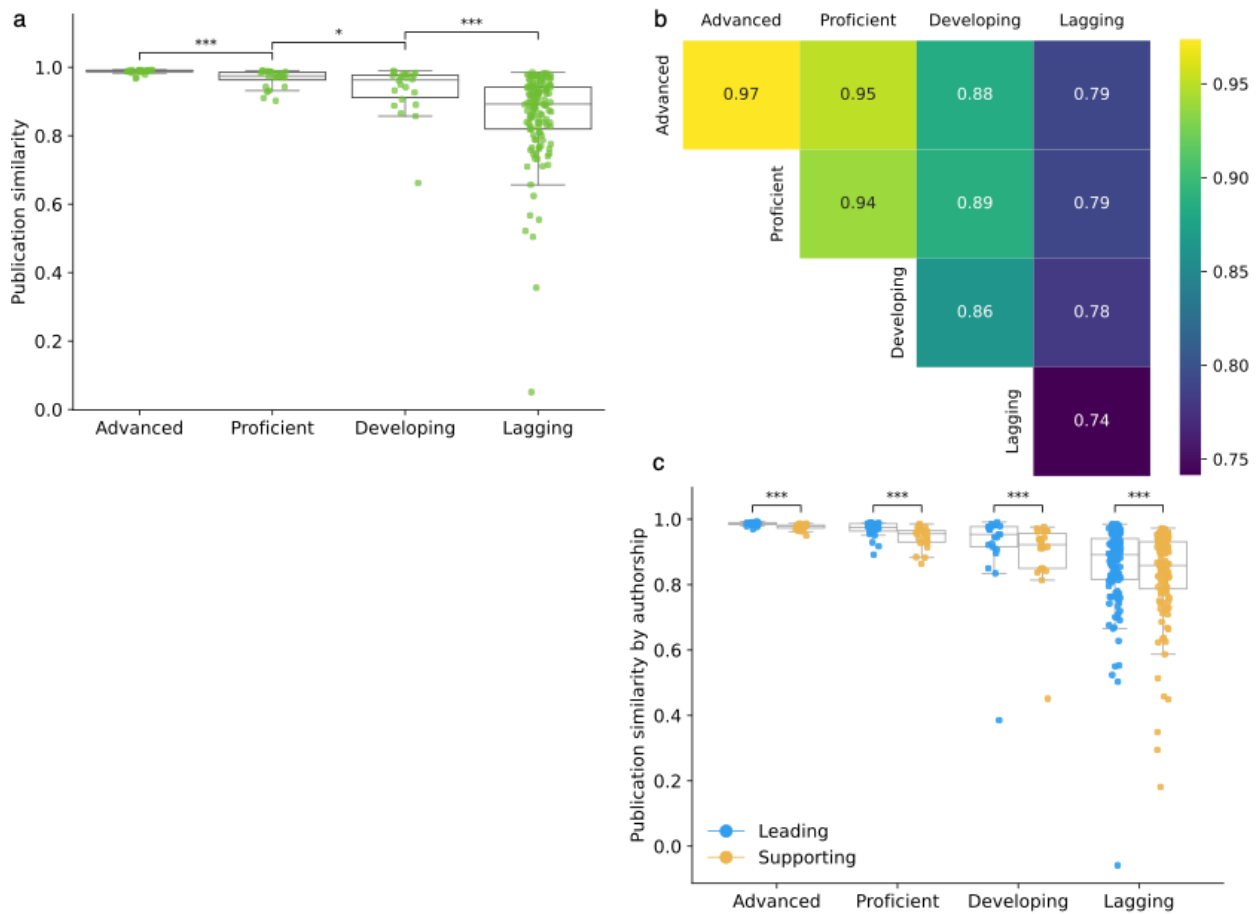

*Figure 5* **The similarity between internationally coauthored publications and domestic publications decreases as the scientific capacity of the country decreases**. *(a) The similarity between internationally coauthored publications and domestic publications within each country, grouped by the country's scientific capacity level. The significance is calculated using a one-sided t-test for independent samples to decide whether the mean value of the group with higher scientific capacity is significantly greater than that of the group with lower scientific capacity. The significance levels are denoted as follows: + indicates P<0.1, ∗ indicates P<0.05, ∗∗ indicates P<0.01, and ∗∗∗ indicates P<0.001. (b) Topical similarity of non-internationally collaborated publications between country pairs, aggregated by scientific capacity level. (c) The same similarity while distinguishing whether authors from the focal country play leading roles or supporting roles. The significance is calculated using a one-sided t-test for related samples to decide whether the mean value of the leading group is significantly greater than the supporting group. The significance levels are denoted as follows: + indicates P<0.1, ∗ indicates P<0.05, ∗∗ indicates P<0.01, and ∗∗∗ indicates P<0.001.*

**Discussion**

Researchers from less-developed countries have long raised concerns about the unequal labor division and the underlying power dynamics in international collaborations. However, comprehensive empirical analyses of this labor division and its discriminatory effects on authors from less-developed countries has been lacking. By examining publication across all disciplines over 200 countries, we provide a large-scale analysis of the relationship between author's country of affiliation and the role in international collaboration. Our findings reveal that labor division in international collaboration follows a hierarchical structure. Researchers from scientifically proficient and developing countries are more likely to assume the role of first author, typically associated with conducting research. In contrast, those from scientifically lagging countries tend to occupy middle-author positions, often linked to supporting research. Meanwhile, researchers from scientifically advanced countries are more likely to take on the last-author role, often functioning as the "brain" of the research team. Notably, these patterns persist even after controlling for influential factors such as authors' scientific capacity and funding status, reaffirming the existence of labor division based on country of affiliation. Furthermore, by examining the topical divergence between domestic and international publications, we reveal that researchers from scientifically lagging countries tend to work on relative different topics when engaging in international collaborations. However, the topics become more aligned with domestic science when these researchers take on leading role in the international collaboration.

Contrary to the ideal of equitable global partnerships, our results suggest that international collaborations are embedded in systemic inequalities. Researchers from less-advanced countries are often relegated to less significant positions, and their research priorities are marginalized. Since authorship is crucial for accumulating scientific capital and reputation, with first and last authors typically receiving the most credit, our findings indicate that researchers from scientifically lagging countries are not only relegated to a subordinate role but are disadvantaged in gaining recognition within international collaborations. Although uneven funding distribution is often cited as a source of power asymmetry, our results show that funding is not the only factor at play. Unfortunately, even funded researchers from lagging countries are less likely to secure the role of last author compared to unfunded researchers from advanced countries.

There are several limitations in our analysis. First and foremost, there are substantial differences in authorship attribution practices across disciplines and countries. Disciplines like arts, humanities and social sciences adhere to a more classical notion of authorship, where writing is the primary contribution garnering authorship and other types of work are often unrewarded[18,26]. In contrast, in biology, the first author is typically attributed to the person who performed the majority of the experiments, and the last author is attributed to the head of the laboratory[26]. Although the authorship practice largely depends on the specific norms of a scientific discipline, the pattern of dominant authors (first and last authors) being more likely associated with conceptual tasks and middle authors being associated with technical tasks hold across the entire spectrum of subfields, even in the social science[26]. Therefore, varying authorship practice does not diminish the validity of our findings that researchers from lagging countries are relegated to the supportive role and tasks in collaborations. Meanwhile, another issue with using authorship to infer the collaboration dynamic is that authorship may not always accurately reflect the true contributions of authors, particularly in the cases of 'ghost authorship' and 'guest authorship'. Individual who contributes significantly to the research may be excluded from authorship, while those included as authors may not have made substantial contributions. Although these unethical practices occur, empirical studies indicate that they are not prevalent: analyses of a sample of medical publications found around 19% publications had evidence of guest authors, and around 11% of publications had evidence of ghost authors[31,32]. Given the relative low proportion of papers with unethical authorship practices, we believe our results are not influenced by such occurrences.

Despite these limitations, our empirical results provide a useful perspective to understand the power asymmetry in current international cooperative frameworks and its adverse effects. This power asymmetry has limited researchers from less-developed countries in fully participating in and benefit from international collaborations, ultimately harming the global scientific community[3] and widening the scientific inequality across countries[9,11]. As science addresses increasingly complex questions, valuing and integrating knowledge from diverse groups is crucial for developing effective and equitable solutions[3,9,33,34]. Furthermore, similar power hegemony can be found across a wide range of conditions, such as setting international trade rules, cooperation and negations on crucial issues like climate change, and strict enforcement of intellectual property rights. This prevalent hegemony has historically maintained the dominance of power countries, often at the expense of the sovereignty and self-determination of weaker countries. We hope our study can provide empirical evidence of this neo-colonialism and serve as a starting point for the global community to develop a more equitable and inclusive framework for cooperation.

**Data and Methods**

**WoS** The dataset is drawn from the Clarivate Analytics' Web of Science database hosted and managed by the Observatoire des Sciences et des Technologies at the University of Montreal. The Web of Science database contains three main citation indices: The Science Citation Index Expanded, the Social Science Citation Index, and the Arts and Humanities Citation Index. We use all indexed papers, including journal articles and review articles, that were published between 2008 to 2020 which in total contains 19,865,673 papers across 201 countries. After excluding publications with missing information in authorship details, author affiliations and demographic information, the final dataset consists of 17,380,209 publications, which accounted for 86% of the total publications.

The nationality of authors is inferred from their affiliations. For authors with multiple affiliations, we select the affiliation with the highest rank. Demographic information of authors included in our analysis contains the gender of authors, the total number of publications indexed in WoS database up to 2022, and the first year the author had a publication in WoS database. Authors are disambiguated using the algorithm developed by Caron and van Eck[35]. The gender of author is inferred through their first name[36].

As we utilize the authorship order to infer the collaborative dynamics in international collaboration, publications ordered alphabetically are removed from the analysis. Among the 3,661,174 internationally coauthored publications, 465,185 papers (13% of internationally coauthored publications) have authors ordered alphabetically. For the analysis involving authorship order, only non-alphabetically ordered publications are included. Although it is possible that the last names of authors may have been accidentally ordered alphabetically in the byline, we adhere to the strict criteria to maintain a clean dataset, ensuring that any publications ordered alphabetically were excluded from the analysis.

Information on the funding of a paper was retrieved from the 'Funding Agency' and 'Grant Number' fields in the WoS. We rely on a previously curated dataset containing the country location information of funding agencies to determine the funding source of a publication[37]. Since funding information is retrieved at the paper level, it is infeasible to link the source of funding with specific authors. Therefore, to estimate whether an author provided funding to the paper, we use an approximate method by examining whether the funding agency is located in the same country as the author. Specifically, if at least one funding agency in the paper comes from the same country as the author, we classify the author as funded.

Countries are classified based on their scientific capacity as developed by Wagner et al[38]. In this classification system, 22 countries are classified as scientifically advanced countries, 24 countries are classified as scientifically proficient countries, 24 countries are classified as scientifically developing countries in the original classification. However, because Yugoslavia and Hong Kong are no longer existing entities, 22 countries are classified as scientifically development countries in our analysis. And the remaining 133 countries are classified as scientifically lagging countries.

**PLOS** To better examine the causal relationship among nationality, author contributions, and authorship positions, we replicate our analysis using a dataset from PLOS journals covering the 2017-2018 period (*N=30770* papers)[29]. This dataset includes detailed author contributions described by the Contributor Roles Taxonomy (CRediT). Since our focus is on the division of scientific labor in international collaborations between researchers from countries with different scientific capacities, we excluded publications with incomplete author demographic information, as well as national publications and international publications coauthored by countries with the same scientific capacity. This resulted in a final dataset of 4,200 papers.

**Authors distribution across scientific capacity groups.** To calculate the number and proportion of authors involved in international collaborations within each scientific capacity group, we assign authors to individual countries based on their affiliation information. Authors with multiple affiliations are attributed to the country of their first listed affiliation. If an author has multiple first affiliations located at different countries within the same year, the author is counted in each corresponding country. International collaborated publications are those where authors, after being assigned to specific countries, come from different countries. Authors of these publications are considered to be participating in international collaboration. After assigning authors to specific countries, we aggregated the counts based on the scientific capacity classification of each country. The proportion in each group is calculated by dividing the number of authors in each classification group by the total number of authors engaged in international collaboration for that year.

**Authorship occurrence normalization.** To calculate the expected value of a specific authorship within countries, we perform 20 random shuffling of the authorship order within each internationally coauthored publication. The normalized authorship occurrence is then calculated using the formula:

$$T_{c,i} = \frac{W_{c,i}}{E_{c,i}}$$

Where $W_{c,i}$ denotes the frequency of researchers from country $c$ appearing in the authorship $i$, $E_{c,i}$ represents the corresponding value obtained from random shuffling, and $T_{c,i}$ is the normalized occurrence of researchers from country $c$ in authorship $i$. We apply the logarithm to $T_{c,i}$. A value large than 0 suggests researchers appear in the authorship more frequently than expected, whereas a value bellow 0 implies researchers from country $c$ are less frequent in authorship $i$.

We apply a similar normalization to the PLOS dataset by randomly shuffling the CRediT contributions of authors within each paper, while keeping the number of contributions made by each author unchanged and ensuring that authors were not assigned duplicate contributions.

th

**Paper-level fixed-effect regression model.** In an ideal scenario, authorship order should reflect the relative contributions made by authors[39]. However, in reality, authorship order is determined through a complex process influenced by many additional factors such as gender[30], age[40] and professional rank[40]. Therefore, to better reveal the causal relationship between potential influential factors and the author's contribution in international coauthored publications, we employ a fixed-effect regression model, treating each individual paper as the analysis unit. Drawing upon these empirical findings, our conceptual model posits that author's contribution is reflected in the authorship order while influenced their gender, nationality, and scientific capacity (see Fig. 3a). We use the number of publications produced by each author up to 2022 as a proxy for their scientific capacity. However, due to the lack of detailed data on the actual contributions of each author, the identified causal relationship between potential factors and contributions—proxied by authorship order—is a mixture of the true relationship and the association between those factors and authorship order.

Specifically, the paper-level fixed-effect regression model defines playing a specific authorship role as the binary outcome while adjusting for paper-level heterogeneity according to the following model:

$$y_i = \log\left(\frac{p}{1-p}\right) = \beta_0 + \beta_1 x_{i,1} + \beta_2 x_{i,2} + \cdots + \alpha_i + \varepsilon_i$$

Where $p$ is the probability that an author is playing the role of a specific authorship, $x_{i,j}$ represents independent variables, $\alpha_i$ accounts for paper-fixed effect, and $\varepsilon_i$ is the error term.

To better measure authors' contributions, we apply the same model to the PLOS dataset which specifies the contributions made by each author. Since an author's labor contribution cannot be accurately captured by any single CRediT classification in the PLOS dataset, we categorize these classifications into two groups: core and peripheral. Building on existing studies[41,42], core contributions include Methodology, Conceptualization, and Supervision, while the remaining tasks are classified as peripheral contributions. An author is labeled as having performed a core task if she contributed to any task within the core group. We first investigate the relationship between potential influencing factors and whether an author performed core tasks. We then use core task performance as an independent variable to examine its relationship with authorship order.

**Measuring the share of credits by researchers**. To measure an individual researcher's contribution, we compute $G_i = \frac{\sum_{j=1}^{K} a_{ij}}{K}$ where $K$ represents the total number of tasks performed by all authors on a given publication. Here $a_{ij} = 1$ indicates that author $i$ performed task $j$ and $a_{ij} = 0$ otherwise. The value of $G_i$ ranges from $1/K$ to 1, where $1/K$ indicates that author $i$ performed only one task, while a value of 1 means that author $i$ performed all tasks. We then average the contributions of all $N$ authors within the same scientific capacity group to compare the average degree of contribution across different scientific capacity groups in international collaboration.

**Measuring the complementarity of contributions**. To assess the extent of labor division in international collaborations, we measure how complementary the contributions of team members are. For each pair of authors within the same publication, complementarity is defined as

$$C_{i,j} = 1 - \frac{|G_i \cap G_j|}{|G_i \cup G_j|}$$

where $G_i$ represents the tasks performed by author $i$, and $G_j$ represents the tasks performed by author $j$. A high $C_{i,j}$ indicates that there is minimal overlap between the tasks performed by authors $i$ and $j$, meaning they contributed to different aspects of the collaboration. Conversely, a low $C_{i,j}$ implies that the authors performed similar tasks. To investigate labor division across authors from different countries, we compute the complementarity for each author pair within each publication and aggregate the complementarity values based on the scientific capacity of the authors' respective countries.

**Measuring similarity between papers.** The content of a publication is represented by a 64-dimensional vector computed from the SPECTER model using title and abstract information[43]. To estimate the similarity between internationally coauthored publications and non-international coauthored publications, we compute the cosine distance between the mean vectors of publications authored internationally and those authored domestically within each discipline for every country. Furthermore, to identify the role of authorship in shaping international collaborations, we distinguish between publications led by authors and those supported by authors. "Led by authors" publications indicate that authors from the focal country played a leading role in the publication, while "supported by authors" publications refer to those in which authors from the focal country played a supportive role. Considering that the first and last authors typically contribute most to the research and often have the greatest influence on the research direction, publications "led by authors" are defined as those where authors from the focal country are either the first or last author, while "supported by authors" publications are those where authors from the focal country are the middle author. If authors from the focal country assume both leading and supporting roles in the publication, we assign the publication to the "led by authors" group.

To ensure that the observed pattern is not an artifact of the number of publications within each country, we conduct a counterfactual experiment assuming that each country produces the same number of publications in each discipline. First, we exclude disciplines from a country where the number of publications is fewer than 20. For the remaining disciplines, we then sample 100 papers with replacement from the raw publication list. Finally, we compare the content similarity using these sampled 100 papers.

# Supplementary Information

**Analysis with publications across different groups**

To better capture the dynamics of international collaborations between countries with varying scientific capacities, we replicate the analysis using internationally coauthored publications involving researchers from different scientific backgrounds. As shown in Fig. S1, the observed patterns persist: researchers from scientifically proficient and developing countries are more likely to occupy the first authorship position, while those from scientifically advanced countries are less likely to serve as middle authors but are more likely to take on the role of last author.

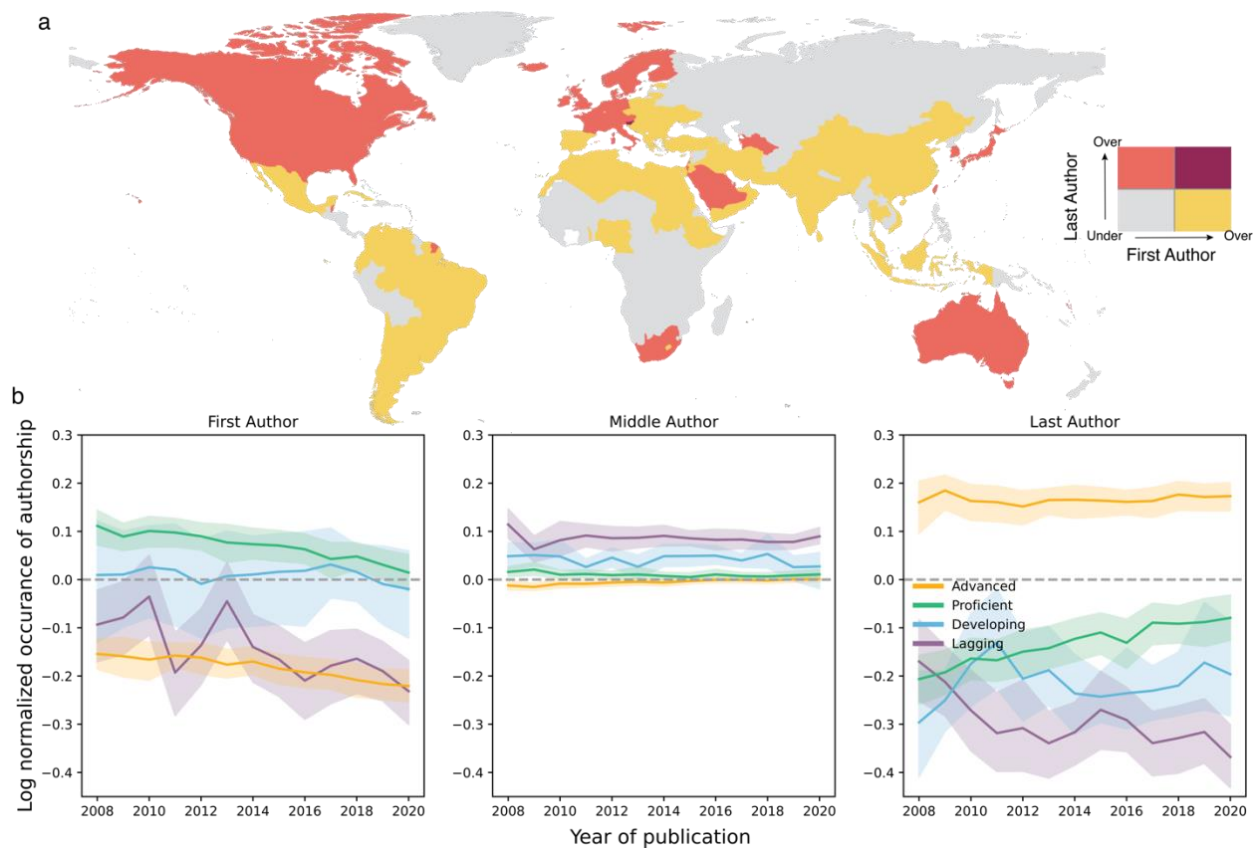

*Figure S1 Authorship representation in publications collaborated by countries with different scientific capacity.*

**Predict authors' contribution with PLOS dataset**

To gain a deeper understanding of labor division in international collaborations, we use a fixed-effects regression model to examine the relationship between key factors and authors' credit assignment.

Table 3 Regression models with CrediT task as dependent variables.

|  | Investigation | Writing - Original Draft | Writing - Review & Editing | Supervision | Formal analysis | Validation | Methodology |
|---|---|---|---|---|---|---|---|
| Male | -0.324*** | -0.258*** | 0.048 | 0.170*** | -0.128*** | -0.107* | -0.124** |
|  | (0.039) | (0.039) | (0.048) | (0.044) | (0.037) | (0.049) | (0.038) |
| Gender(Unknown) | -0.092+ | -0.173** | -0.264*** | -0.114+ | -0.245*** | -0.189** | -0.232*** |
|  | (0.054) | (0.056) | (0.070) | (0.063) | (0.052) | (0.070) | (0.055) |
| Log no.pub | -0.211*** | 0.082*** | 0.540*** | 0.720*** | -0.050*** | 0.126*** | 0.067*** |
|  | (0.013) | (0.011) | (0.015) | (0.018) | (0.011) | (0.015) | (0.012) |
| Proficient | 0.411*** | -0.003 | -0.579*** | 0.220*** | -0.048 | -0.052 | -0.104+ |
|  | (0.056) | (0.052) | (0.068) | (0.058) | (0.055) | (0.069) | (0.056) |
| Developing | 0.652*** | -0.276** | -0.628*** | 0.042 | -0.397*** | -0.370** | -0.603*** |
|  | (0.113) | (0.104) | (0.128) | (0.110) | (0.108) | (0.136) | (0.116) |
| Lagging | 0.396*** | -0.790*** | -0.817*** | 0.325*** | -1.069*** | -0.703*** | -0.961*** |
|  | (0.074) | (0.069) | (0.093) | (0.077) | (0.070) | (0.089) | (0.075) |
| Num.Obs. | 26366 | 29918 | 20803 | 27753 | 28856 | 18111 | 26500 |
| R2 | 0.201 | 0.161 | 0.299 | 0.271 | 0.163 | 0.181 | 0.183 |
| R2 Adj. | 0.021 | -0.052 | 0.119 | 0.071 | -0.028 | -0.016 | 0.002 |

+ p < 0.1, * p < 0.05, ** p < 0.01, *** p < 0.001

|  | Funding acquisition | Project administration | Resources | Data curation | Conceptualization | Visualization | Software |
|---|---|---|---|---|---|---|---|
| Male | 0.059 | -0.056 | 0.211*** | -0.247*** | -0.040 | -0.156** | 0.313*** |
|  | (0.047) | (0.044) | (0.047) | (0.041) | (0.038) | (0.054) | (0.062) |
| Gender(Unknown) | -0.019 | -0.050 | 0.108+ | -0.071 | -0.048 | -0.324*** | 0.257** |
|  | (0.069) | (0.062) | (0.065) | (0.055) | (0.056) | (0.078) | (0.088) |
| Log no.pub | 0.676*** | 0.308*** | 0.319*** | -0.212*** | 0.462*** | -0.039* | -0.145*** |
|  | (0.019) | (0.014) | (0.016) | (0.012) | (0.013) | (0.015) | (0.018) |
| Proficient | 0.448*** | 0.564*** | 0.471*** | 0.277*** | 0.113* | -0.010 | -0.043 |
|  | (0.063) | (0.061) | (0.069) | (0.060) | (0.055) | (0.072) | (0.088) |
| Developing | 0.120 | 0.391*** | 0.469*** | 0.219+ | -0.170 | -0.429** | -0.232 |
|  | (0.132) | (0.108) | (0.138) | (0.117) | (0.110) | (0.147) | (0.167) |
| Lagging | -0.181* | 0.801*** | 0.792*** | 0.100 | -0.498*** | -1.099*** | -0.847*** |
|  | (0.087) | (0.075) | (0.097) | (0.078) | (0.071) | (0.108) | (0.116) |
| Num.Obs. | 25000 | 23948 | 20168 | 25091 | 28743 | 16035 | 12489 |
| R2 | 0.236 | 0.181 | 0.197 | 0.186 | 0.216 | 0.163 | 0.136 |
| R2 Adj. | 0.017 | -0.026 | 0.008 | -0.003 | 0.031 | -0.050 | -0.093 |

+ p < 0.1, * p < 0.05, ** p < 0.01, *** p < 0.001

**Counterfactual scenario assuming the same number of publications**

To ensure that the observed topical divergence within and across countries is not simply due to sampling bias from varying publication numbers, we constructed a counterfactual publication portfolio for each country within each discipline and replicated the analysis (see Data and Methods). The results continue to support an inverse relationship between a country's scientific capacity and its topical divergence between domestic and international publications (see Fig. S2a,c). However, in the counterfactual scenario, the divergence is smaller, and the differences among scientific groups are less pronounced (see Fig. S2a,c). The topical similarity across countries with varying scientific capacities remains consistent with the results derived from the actual data (see Fig. S2b).

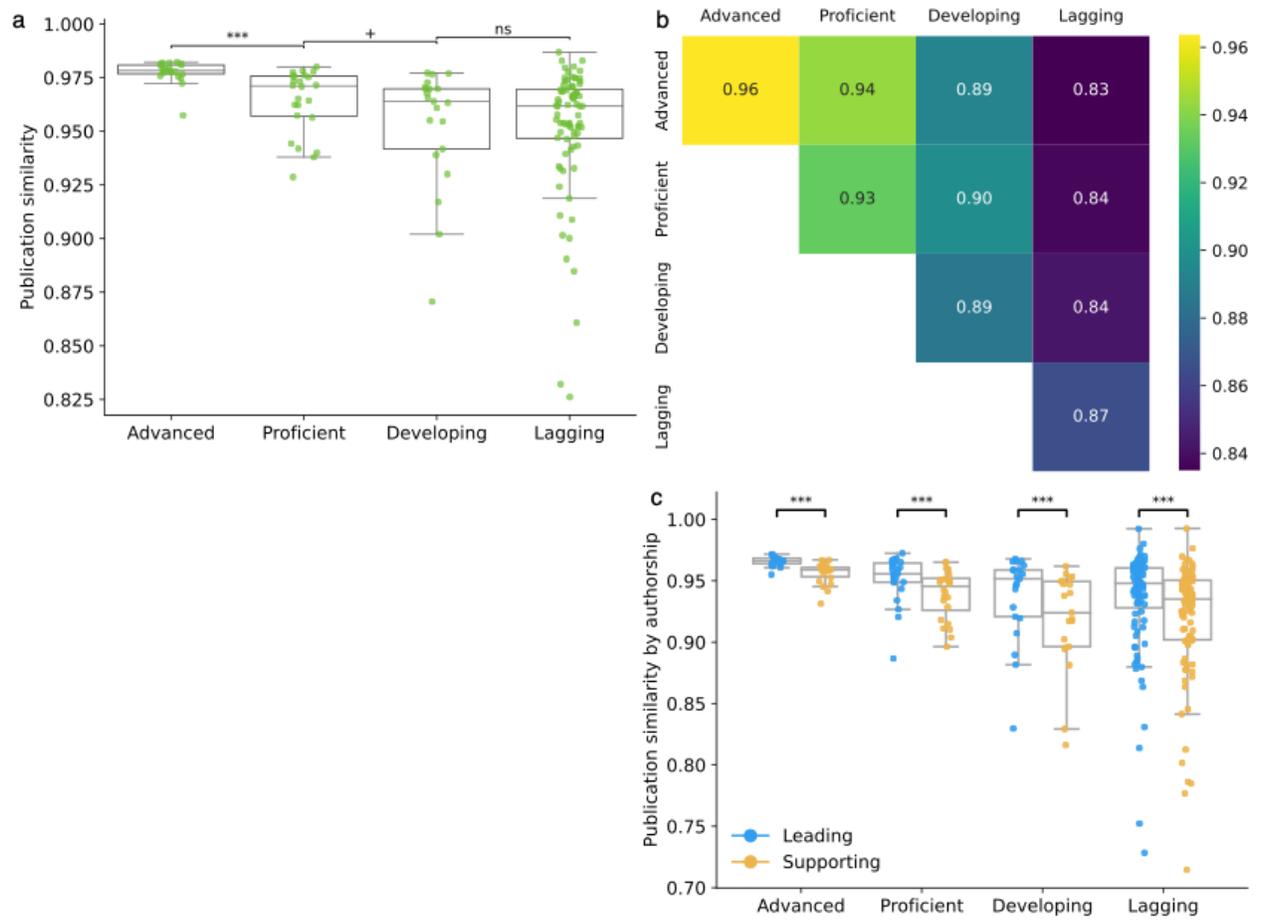

*Figure S2 Topical divergence derived from counterfactual scenario in which each country is assumed to produce the same number of publications in each discipline.*